\documentclass[journal=jacsat,manuscript=article]{achemso}

\usepackage[version=3]{mhchem} 
\usepackage{graphicx,amsmath,amssymb,epsfig,color}
\usepackage{dcolumn}
\usepackage{bm}
\usepackage{CJK}
\author{Junfeng Su}
\affiliation
{State Key Laboratory of Precision Spectroscopy, East China Normal University, Shanghai 200062, China}
\author{Datang Xu}
\affiliation
{State Key Laboratory of Precision Spectroscopy, East China Normal University, Shanghai 200062, China}
\author{Guoxiang Huang}
\affiliation
{State Key Laboratory of Precision Spectroscopy, East China Normal University, Shanghai 200062, China}
\email{gxhuang@phy.ecnu.edu.cn}

\title{Storage and Retrieval of Surface Polaritons}

\abbreviations{IR,NMR,UV}
\begin{document}
\begin{abstract}
We investigate the memory of surface polariton (SP) via the electromagnetically induced transparency (EIT) of quantum emitters doped at the interface between a dielectric and a metamaterial. We show that, due to the strong mode confinement provided by the interface, the EIT effect can be largely enhanced; furthermore, the storage and retrieval of the SP can be realized by switching off and on of a control laser field; additionally, the efficiency and fidelity of the SP memory can be improved much by using a weak microwave field. The results reported here are helpful not only for enhancing the understanding of SP property but also for promising applications in quantum information processing and transmission.
\begin{figure}
  \centering
  \includegraphics[scale=0.5]{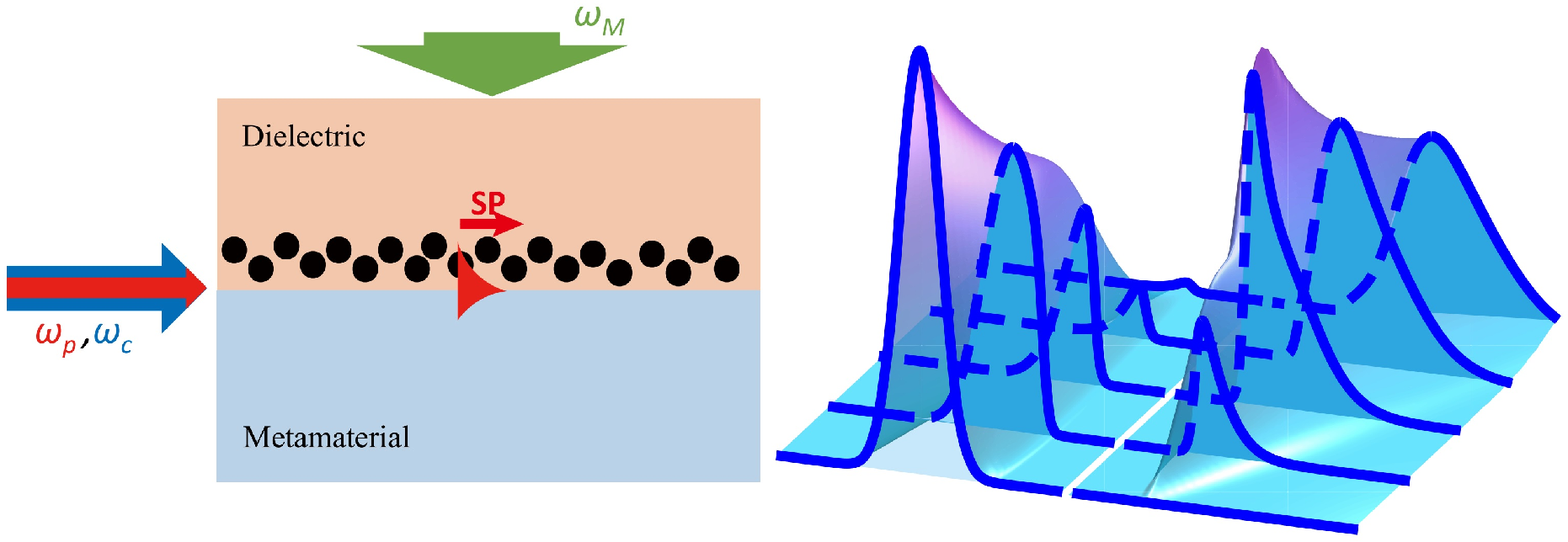}\\
  \label{TOC}
\end{figure}

\end{abstract}





In past two decades, considerable attention has been paid to the research on electromagnetically induced transparency (EIT), a typical quantum interference effect occurring in resonant three-level atomic gases. The light behavior in EIT systems possesses many striking features, including substantial suppression of optical absorption, significant slowdown of group velocity, excellent coherent control of the interaction between light and atoms at weak light level, and so on~\cite{Fle1}.

One of important applications of EIT is light storage, which has important applications in quantum informatics~\cite{Lvo}. The basic mechanism of EIT-based light storage is that the system allows a stable combined excitation (called dark-state polariton~\cite{Fle2}) contributed by atomic coherence together with a probe (or called signal) laser field, which displays an atomic character when a control laser field is switched off and a photonic character when the control field is switched on. The storage of probe optical pulses based on atomic EIT has been verified in many experiments~\cite{Nov,Bre}. Recently, the possibility of weak-light soliton memory in a cold atomic gas has also been suggested~\cite{Chen}.

Up to now most of works on EIT-based light storage were carried out in free atomic gases. Although there are some advantages (e.g., long coherence time), such EIT and light storage scheme require special and cumbersome conditions, such as large device size, which hamper compact chip-integrated applications. In contrast, solid media is on demand for practical applications because they are easy for device integration. Some materials (like quantum dots or dopant ions in solids) can combine advantages of atomic gases and solids to get long coherence time and large optical density. Thus the realization of the light storage based on such hybrid materials are more desirable comparing with the atomic gases in free space~\cite{Zol,Sim}.

In recent years, much effort has been focused on the study of surface polaritons (SPs), i.e., polarized electromagnetic waves propagating along a metal-dielectric interface with the wave coupled to charge density oscillations~\cite{Maier}. SPs produced in metal-dielectric interfaces doped with quantum emitters have also been considered in many studies~\cite{Stock,Ber,Berg,Chan,Akim,Leon,Nogi,Outl,Dzso,Bol,Dorfman13,zhu14}.
In the work by Kamli {\it et al}.~\cite{Kam1,Kam2}, an interesting scheme was proposed for realizing a coherent control of low-loss SPs by doping 3-level quantum emitters at the interface between a dielectric and a negative index metamaterial (NIMM) working under EIT condition, which hints the possibility to realize light memory by using plasmonic metamaterials. A scheme to obtain a stable propagation of nonlinear SPs by placing a $N$-type 4-level quantum emitters at a NIMM-dielectric interface was also suggested recently~\cite{Tan}.

In the present work, we investigate, both analytical and numerically, the SP memory in $\Lambda$-type three-level quantum emitters doped at a dielectric-NIMM interface. We shall demonstrate that, due to the strong transverse mode confinement contributed by the interface, the EIT effect can be largely enhanced. Furthermore, the storage and retrieval of the SP propagating along the interface can be realized by switching off and on of a control laser field. In addition, the efficiency and fidelity of the SP memory can be improved much by using a weak microwave field, which is applied within the time interval between the switching off and on of the control field. The results reported here are useful not only for a deep understanding of the physical property of SPs but also for practical applications in light and quantum information processing and transmission based on hybrid plasmonic metamaterials.

We stress that, comparing with the light storage by using atomic gases in free space, the SP memory suggested here possesses many advantages, including: (i)~The quantum emitters are fixed in space, the incoherence induced by the atomic diffusion is avoided; (ii)~The density of the quantum emitters is higher than that of free atomic gases and light field in the system is tightly confined at the NIMM-dielectric interface, which are useful for obtaining a strong coupling between the light field and the quantum emitters, and hence easy for realizing effective control and storage of optical pulses; (iii)~The size of the device can be made small, which is promising for compact chip-integrated applications.

\section{MODEL}

We consider an interface consisting of a NIMM as bottom plane (with permittivity $\varepsilon_1$ and permeability $\mu_1$) and a dielectric as top plane (with permittivity $\varepsilon_2$ and permeability $\mu_2$). Quantum emitters with $\Lambda$-type 3-level configuration (denoted by solid black dots, which in general can be atoms, quantum dots, nitrogen-valence centers, rare-earth ions, etc.~\cite{Kam1}) are dropped into a thin layer of the dielectric near the NIMM-dielectric interface (see Fig.~\ref{fig1}). The probe and control fields (with angular frequencies $\omega_p$ and $\omega_c$, respectively) are incident along the interface (i.e. $x$ direction); the microwave field (with angular frequency $\omega_M$) is incident from $-z$ direction.
Fig.~{\ref{fig1}(a) shows energy states $|j\rangle$ $(j=1,2,3)$  and the excitation scheme of the quantum emitters, with $\Delta_3$ and $\Delta_2$ respectively the one-photon and two-photon detunings.
Due to the coupling between the light fields and the NIMM-dielectric interface, SP can be produced and propagate along the $x$ direction. To suppress the noise induced by four-wave mixing,
we assume the initial population of the quantum emitters is prepared in the
metastable state $|1\rangle$~\cite{Vur}.

\begin{figure}
  \centering
  \includegraphics[scale=0.6]{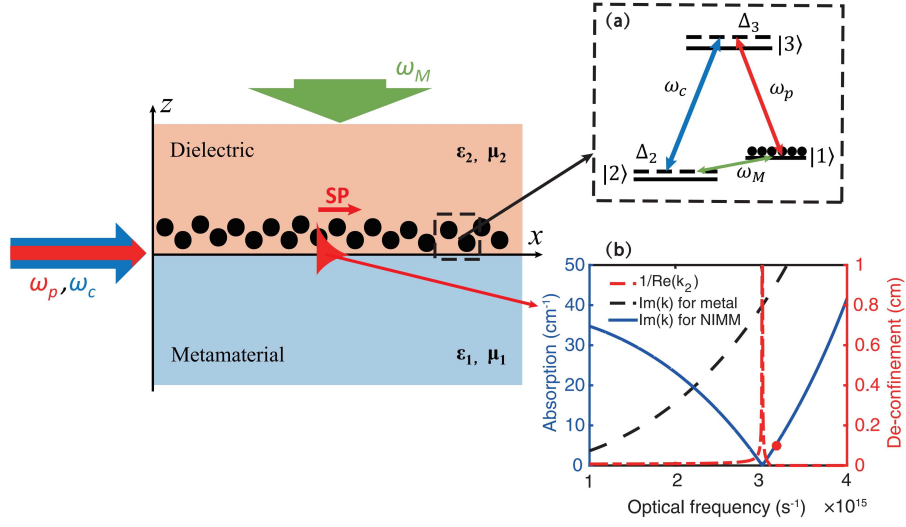}\\
  \caption{\small (color online) Surface polariton (SP) excited at the interface of the NIMM ($\varepsilon_1$, $\mu_1$) and the dielectric ($\varepsilon_2$, $\mu_2$), propagating along the interface. The probe and control fields (with angular frequencies $\omega_p$ and $\omega_c$, respectively) are incident along the interface (i.e. $x$ direction). The microwave field (with angular frequency $\omega_M$) is incident from $-z$ direction.  Inset: (a) Levels and excitation scheme of the $\Lambda$-type 3-level quantum emitters (denoted by solid black dots); $\Delta_3$ ($\Delta_2$) is the one-photon (two-photon) detuning. (b) Absorption $\textrm{Im}(k)$ of the SP for the NIMM-dielectric interface (blue solid line) and for a metal-dielectric interface (black dashed line), and de-confinement $1/\textrm{Re}(k_2)$ (red dashed line) as functions of optical oscillating frequency $\omega_l$. For suppressing FWM noise the initial population is prepared at the metastable state $|1\rangle$.}\label{fig1}
\end{figure}

Different from metal-dielectric interface, the NIMM-dielectric interface supports both TE modes and TM modes. Here we consider the TM mode only, which, by solving the Maxwell equations with boundary conditions, has the form
\begin{equation}\label{eq1}
  \textbf{E}(\textbf{r},t)=\left\{\begin{array}{ll}
(k\textbf{e}_z+ik_1\textbf{e}_x)\frac{c}{\varepsilon_1\omega_l}
\sqrt{\frac{\hbar\omega_l}{\varepsilon_0L_xL_yL_z}}a(\omega_l)
e^{k_1z+i(kx-\omega_lt)}+c.c.&\text{($z<0$)},\\
(k\textbf{e}_z-ik_2\textbf{e}_x)\frac{c}{\varepsilon_2\omega_l}
\sqrt{\frac{\hbar\omega_l}{\varepsilon_0L_xL_yL_z}}a(\omega_l)
e^{-k_2z+i(kx-\omega_lt)}+c.c.&\text{($z>0$)},
\end{array}\right.
\end{equation}
where ${\bf r}=(x,z)$  (for simplicity  we assume the incident electric field has a large width in $y$ direction so that
all quantities are independent of $y$),
$\textbf{e}_x$ ($\textbf{e}_z$) is the unit vector along the $x$ ($z$) direction, $\omega_l$ is oscillating frequency, $k_{\alpha}^2=k^2-\omega_l^2\varepsilon_{\alpha}\mu_{\alpha}$ ($\alpha=1,2$) is the wave vector satisfying the relation $k_1\varepsilon_2=-k_2\varepsilon_1$, $k=(\omega_l/c)[\varepsilon_1\varepsilon_2(\varepsilon_1\mu_2-\varepsilon_2\mu_1)/(\varepsilon_1^2-\varepsilon^2_2)]^{1/2}$ is the propagation constant of the SP mode in the absent of the quantum emitters, $L_x$ and $L_y$ are respectively lengthes of the NIMM-dielectric interface along $x$ and $y$ directions, and $L_z$ is the effective length of $z$ direction (see Sec.~A of the Supporting information). Here we assume that the photon number of all laser fields are much larger than unity and hence
the quantity $a(\omega_l)$ can be taken as a c-number.

The NIMM can be described by using macro-parameters, i.e. the permittivity and the permeability, which are given
by~\cite{Kam1,Kam2,Tan}  $\varepsilon_1(\omega_l)=\varepsilon_\infty-\omega^2_e/[\omega_l(\omega_l+i\gamma_e)]$ and $\mu_1(\omega_l)=\mu_\infty-\omega^2_m/[\omega_l(\omega_l+i\gamma_m)]$, where $\omega_{e,m}$ are electric and magnetic plasma frequencies of the NIMM, $\gamma_{e,m}$ are corresponding decay rates, $\varepsilon_\infty$ and $\mu_\infty$ are background constants. Fig.~\ref{fig1}(b) shows the imaginary part of the propagation constant $k$ of the SP (blue solid line)
as a function of the frequency $\omega_l$. We see that
comparing with conventional metal-dielectric interface (black dashed line), there is a high suppression of SP loss along the NIMM-dielectric interface. This is due to the destructive interference between electric and magnetic responses in the NIMM~\cite{Kam1,Kam2,Tan}. When plotting the figure, the parameters are chosen as $\varepsilon_\infty=\mu_\infty=1, \omega_e=1.37\times10^{16}\textrm{s}^{-1}, \gamma_e=2.73\times10^{13}\textrm{s}^{-1}$(as for Ag), $\omega_m=3.21\times10^{15}\textrm{s}^{-1}, \gamma_m=10^{11}\textrm{s}^{-1}, \varepsilon_2=1.3$, and $\mu_2=1$.

From the figure, it seems that there is a possibility to achieve a complete suppression of the SP loss. However,
such complete suppression of the SP loss is unavoidably accompanied by a de-confinement of the SP in the dielectric, i.e., when Im$(k)\rightarrow0$, the de-confinement parameter describing the spatial decay of the electric field in $+z$ direction, i.e. 1/Re$(k_2)$, grows rapidly  [see the red dashed line of Fig.~\ref{fig1}(b)\,]. Thus one must
select an appropriate excitation frequency with a small deviation from the lossless point (e.g. the red solid circle, which corresponds to $\omega_l=3.11\times 10^{15}$\,Hz), where the SP loss is still suppressed greatly and a strong SP confinement can be also achieved simultaneously.

\section{RESULTS AND DISCUSSION}


{\large \bf EIT enhancement}.  The SP mode (\ref{eq1}) will be modified when the quantum emitters are included. In this case the system is a hybrid one, and the SP mode takes still the form of (\ref{eq1}) but the quantity $a$ will be modulated in space and time, i.e. $a=a(\omega_l, \textbf{r},t)$. Assuming that the probe and control fields are coupled to the SP mode, we have $\textbf{E}(\textbf{r},t)=\Sigma_{l=p,c}\textbf{u}_l(z)\xi_l (\textbf{r},t)\exp\{i[k(\omega_l)x-\omega_lt]\}+{\rm c.c.}$,  with $\xi_l (\textbf{r},t)=[\hbar\omega_l/(\varepsilon_0L_xL_yL_z)]^{1/2}a(\omega_l, \textbf{r},t)$ and $\textbf{u}_l(z)=c[k(\omega_l)\textbf{e}_z+ik_2(\omega_l)\textbf{e}_x]\exp[k_2(\omega_l)z]/(\varepsilon_2\omega_l)$; the microwave field is not coupled to the SP mode and is assumed to be a continuous wave.
The dynamics of the system in interaction picture is governed by the optical Bloch equation
\begin{equation}\label{Bloch}
  i\hbar\left(\frac{\partial}{\partial t}+\Gamma\right)\sigma=[H_{\textrm{int}},\sigma],
\end{equation}
where $\sigma$ is a $3\times 3$ density matrix and $\Gamma$ is a $3\times 3$ relaxation matrix (see the \textit{Supporting information} for detail). The interaction Hamiltonian reads
$H_{\textrm{int}}=-\hbar[\sum^3_{j=1}\Delta_{j}|j\rangle\langle j|+(\zeta_p(z)\Omega_p|3\rangle\langle 1|+\zeta_c(z)\Omega_c|3\rangle\langle 2|+\Omega_m|2\rangle\langle 1|+{\rm h.c.})]$,
where $\Delta_1=0$, $\Delta_2=\omega_p-\omega_c-\omega_{21}$ and $\Delta_3=\omega_p-\omega_{31}$ ($\omega_{jl}$ is the difference
of the eigenfrequencies between the state $|j\rangle$ and the state $|l\rangle$);
$\Omega_p (\textbf{r},t)=\xi_p (\textbf{r},t)|\textbf{p}_{31}|/\hbar$, $\Omega_c =\xi_c|\textbf{p}_{32}|/\hbar$, and $\Omega_m=E_m(x)|\textbf{p}_{21}|/\hbar$ are respectively the half Rabi frequencies of the probe, control and microwave fields; $\textbf{p}_{jl}$ is the electric dipole matrix element associated with the transition from the state $|j\rangle$ to the state $|l\rangle$. Note that here $\xi_c$ is assumed to a constant since the control field is strong and thus can be taken to be undepleted during the evolution of the probe field.
In addition, one can take $\zeta_c(z)\approx\zeta_p(z)=|\textbf{u}_p(z)|\equiv\zeta(z)$ in the calculation since $\omega_p\approx\omega_c$.  Since the mode function $\textbf{u}_p(z)$ (and hence $\zeta(z)$) is a fast varying function of $z$,
the Bloch Eq.~(\ref{Bloch}) can be simplified into a reduced one by a transformation $\sigma_{jl}(x,z,t)\rightarrow \tilde{\sigma}_{jl}(x,t)$. See Sec.~B of the Supporting information.

The evolution of the probe field is described by the Maxwell equation

\begin{equation}\label{Maxwell}
  i\left(\frac{\partial}{\partial x}+\frac{n^2_2}{n_{\textrm{eff}}c}\frac{\partial}{\partial t}\right)\Omega_p+\langle\kappa_{13}(z)\widetilde{\sigma}_{31}\rangle=0,
\end{equation}
where $n_2$ is the refractive index of the dielectric, $\kappa_{13}(z)\equiv N_{QE}(z)|\textbf{p}_{13}|\omega_p/(2n_{\textrm{eff}}\hbar\varepsilon_0c)$, with $N_{QE}(z)$ the emitter density and $n_{\textrm{eff}}(\equiv ck/\omega_p$) an effective refractive index.
Because the quantum emitters are doped in a thin layer near the NIMM-dielectric interface, we assume
$N_{QE}(z)= N_a$ for $0<z<z_0$ and 0 for $z<0$ and $z>z_0$, where $N_a$ is a constant and $z_0(=2\mu$m) is the thickness of the layer.
In addition, in Eq.~(\ref{Maxwell}) we have defined   $\sigma_{31}=\zeta(z)\widetilde{\sigma}_{31}$ and $\langle\psi\rangle=\int^{+\infty}_{-\infty}dz|\zeta(z)|^2\psi(z)/(\int^{+\infty}_{-\infty}dz|\zeta(z)|^2)$, here $\psi$ is any function (see Sec.~B of the Supporting information).

We first study the time evolution of the system when the microwave field is absent, which can be obtained by
linearizing the Maxwell-Bloch (MB) equations (~\ref{Bloch}) and (~\ref{Maxwell}) around the initial state
($\sigma^{(0)}_{11}=1$ and other $\sigma_{ij}^{(0)}=0$). We obtain
$\Omega_p=F\,e^{i[K(\omega)x-\omega t]}$, $\sigma_{21}=a^{(1)}_{21}\zeta(z)F\,e^{i[K(\omega)x-\omega t]}$ and $\sigma_{31}=a^{(1)}_{31}\zeta(z)F\,e^{i[K(\omega)x-\omega t]}$, with other $\sigma_{jk}=0$. Here $F$ is a constant and $K$ is the linear dispersion relation of the SP
\begin{equation}\label{dispersion}
  K(\omega)=\frac{n_2\omega}{n_{\textrm{eff}}\,c}+\left\langle\frac{\kappa_{13}(z)(\omega
  +d_{21})}{|\zeta(z)\Omega_c|^2-(\omega+d_{21})(\omega+d_{31})}\right\rangle.
\end{equation}
The imaginary and real parts of $K$, i.e. Im($K$) and Re($K$), as functions of $\omega$ (i.e. the deviation from the center frequency $\omega_p$) for the case of the SP excitation are illustrated by the blue solid lines in Fig.~\ref{fig2}(a) and Fig.~\ref{fig2}(b), respectively.
%
\begin{figure}
\centering
\includegraphics[scale=0.55]{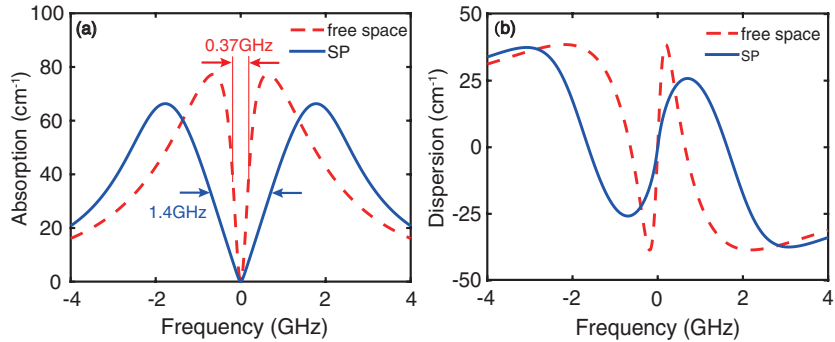}\\
\caption{\small (color online) Linear dispersion relation of the SP excited in the NIMM-dielectric interface doped with quantum emitters. (a) Absorption Im($K$) and (b) dispersion Re($K$) as functions of $\omega$. Blue solid lines (red dashed lines) are for the case of the SP excitation here and for the case of conventional EIT in free atomic gas, respectively.
}\label{fig2}
\end{figure}
%
For comparison, corresponding quantities for the case of conventional EIT of an atomic gas in free space are also illustrated by the red dashed lines.

When plotting the figure, the both control fields are assumed to have the same energy for the two cases, which gives $\Omega^{\textrm{\rm sp}}_c/\Omega^{\rm free}_c=3.27$, where $\Omega^{\rm sp}_c$ ($\Omega^{\textrm{\rm free}}$) is the half Rabi frequency of the control field in the present SP system (the atomic gas in free space)~\cite{note100}; the thickness of the doped layer is taken to be 2\,$\mu$m and light beams be Gaussian with beam waist 10\,$\mu$m.
Parameters of the quantum emitter are  $\omega_p=3.11\times10^{15}$\,Hz, $\Gamma_2=0.02\,\textrm{Hz}$, $\Gamma_3=12\,\textrm{kHz}$, $\gamma_{21}=10\,\textrm{kHz}$, $\gamma_{31}=\gamma_{32}=9\,\textrm{kHz}$, $\Delta_2=\Delta_3=0$, $\Omega_c=2\,\textrm{GHz}$, $N_a=4.7\times10^{18}\textrm{cm}^{-1}$ and $\kappa_{13}(0)=1.55\times10^{11}\textrm{s}^{-1}\textrm{cm}^{-1}$.
Note that the inhomogeneous broadening at transitions $|1\rangle\rightarrow|3\rangle$ and $|1\rangle\rightarrow|2\rangle$ have generally Gaussian lineshapes, but for simplicity here they are approximated as Lorentzian ones~\cite{Kuznetsova}
with the forms $W_{31(21)}/[\pi(W^2_{31(21)}+\triangle\omega^2_{31(21)})]$. Here $W_{31}=2\,\textrm{GHz}$ and $W_{21}=30\,\textrm{kHz}$ are widths of the inhomogeneous broadening and $\triangle\omega_{31(21)}$ are corresponding energy level shift, corresponding respectively to the transitions $|1\rangle\rightarrow|3\rangle$ and $|1\rangle\rightarrow|2\rangle$.

From Fig.~\ref{fig2}(a), we see that the width of the EIT transparency window of the probe field is about 4 times larger than that in the free atomic gas; furthermore, height of the two absorption peaks are also smaller than case of the free atomic gas. In addition, from Fig.~\ref{fig2}(b) we see that near $\omega=0$ (corresponding to the center frequency of the probe field), the slope of Re($K$) (and hence the group velocity of the SP field) is much smaller than the case of the free atomic gas.  All these interesting characters appeared in the SP field is a reflection of the enhancement of EIT effect, which are mainly due to the mode confinement contributed by the NIMM-dielectric interface, and helpful for
the light storage and retrieval in the system.

For a Gaussian input of the probe field, i.e. $\Omega_p(0,t)=\Omega_p(0,0)\textrm{exp}(-t^2/\tau^2_0)$, by solving
Eq.~(\ref{Maxwell}) we obtain the Rabi frequency of the SP pulse with the form
\begin{equation}\label{linearsolu}
\Omega_p(x,t)=\frac{\Omega_p(0,0)}{\sqrt{b_1(x)-ib_2(x)}}\exp\left\{iK_0x-\frac{(K_1x-t)^2}{[b_1(x)-ib_2(x)]\tau^2_0}\right\},
\end{equation}
where $K_j=\partial^j K/\partial \omega^j|_{\omega=0}$, $b_1(x)=1+2x\textrm{Im}(K_2)/\tau^2_0$ and $b_2(x)=2x\textrm{Re}(K_2)/\tau^2_0$. Corresponding electric field reads
\begin{equation}\label{electricfield}
  \textbf{E}_p(\textbf{r},t)=\frac{\hbar \Omega_p(0,0)\textbf{u}_p(z)}{|\textbf{p}_{13}|\sqrt{b_1(x)-ib_2(x)}}
  \exp\left\{ik(\omega_p)x+iK_0x-\frac{(K_1x-t)^2}{[b_1(x)-ib_2(x)]\tau^2_0}-i\omega_pt\right\}+{\rm c.c.}
\end{equation}
With $\Omega_c=300\textrm{MHz}$, $\Delta_2=1\times10^6$\,Hz, $\Delta_3=5\times10^8$\,Hz and the other parameters the same as those given above,
we obtain $\textrm{Re}(k)=1.1822\times10^5\,\textrm{cm}^{-1}$, $\textrm{Im}(k)=3.0929\,\textrm{cm}^{-1}$, $K_0=(3.0791+0.2718i)\,\textrm{cm}^{-1}$, $K_1=(3.0963+0.3496i)\times10^{-6}\,\textrm{cm}^{-1}\textrm{s}$, and $K_2=(0.0839+3.5700i)$ $\times10^{-13}\,\textrm{cm}^{-1}\textrm{s}^2$. The group velocity of the SP reads
$V_g=1/K_1=1.0766\times10^{-5}c$,
i.e. the SP moves very slowly along the NIMM-dielectric interface comparing with light speed in vacuum.
If the probe pulse duration $\tau_0=1.5\times10^{-7}$\,s, we obtain the spatial length $l_{\textrm{sp}}=V_g\tau_0=0.48\textrm{mm}$
of the SP, which means that there is a large compression of the SP pulse and it is helpful for the SP memory in the system.

{\large \bf Storage and retrieval of SPs without microwave field}. We now investigate the SP memory in the absence of the microwave field. Such task can be accomplished by the switching-off and switching on of the control field, which can be modeled by the combination of two hyperbolic tangent functions with the form~\cite{Chen}
\begin{equation}\label{switch1}
\Omega_c(t)=\Omega_{c0}\left[1-\frac{1}{2}\tanh\left(\frac{t-T_1}{T_s}\right)
+\frac{1}{2}\tanh\left(\frac{t-T_2}{T_s}\right)\right],
\end{equation}
where $T_1$ and $T_2$ are respectively times of switching-off and switching-on of the control field.  The switching time of the control field is $T_{s}$ and the storage time of the SP is approximately given by $T_2-T_1$.

The principle of EIT-based SP memory is as follows. When the control field is applied, the SP pulse propagates along the NIMM-dielectric interface; by switching off the control field at $t=T_1$, the propagating velocity of the SP approaches zero and the SP disappears and gets stored in the medium in the form of the coherence of the quantum emitters; when the control field is switched on at $t=T_2$ the SP appears again. Shown in Fig.~\ref{fig3}(a)
\begin{figure}
\centering
\includegraphics[scale=0.41]{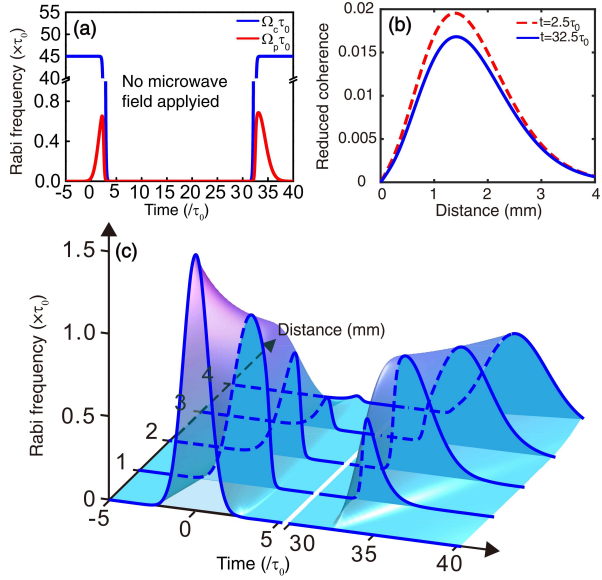}
\caption{\small (color online) SP memory in the absence of microwave field. (a)~Dimensionless temporal profile of the control field $\Omega_c\tau_0$ (blue solid line)  and the SP field $\Omega_p\tau_0$ (red solid line) at position  $x=2\,\textrm{mm}$. (b)~The coherence of the quantum emitters, $\tilde{\sigma}_{21}$, as a function of distance $x$ at the time $t_1=2.5\, \tau_0$ (red dashed line) and $t_2=32.5\,\tau_0$ (blue solid line). (c)~The SP  waveshape as a function of $x$ and $t$ during the storage and retrieval in the whole medium.}
\label{fig3}
\end{figure}
is the numerical result of the temporal profile of the control (blue solid line) and the SP (red solid line) fields for the storage and retrieval at the position $x=2\,\textrm{mm}$, by taking $\Omega_{c0}\tau_0=45$ (i.e. $\Omega_{c0}=300\,$MHz), $T_1=2.5\tau_0$, $T_2=32.5\tau_0$, and $T_s=0.2\tau_0$, with $\tau_0=1.5\times10^{-7}$\,s. We see that the SP can indeed be stored and retrieved in the system.

Fig.~\ref{fig3}(b) shows the coherence of the quantum emitters, $\tilde{\sigma}_{21}$, during the SP memory as a function of $x$ at the time $t_1=2.5\, \tau_0$ (red dashed line) and $t_2=32.5\,\tau_0$ (blue solid line). We see that, during the period of storage, $\tilde{\sigma}_{21}$ is reduced due to the dephasing of the system, which leads to the decrease of the efficiency and fidelity of the SP memory.
Fig.~\ref{fig3}(c) shows the result of the SP waveshape as a function of $x$ and $t$ during the storage and retrieval in the whole medium (with medium length $L_x=4$\,mm), by taking $\Omega_p\,(0,t)\tau_0=1.5\exp[-(t/\tau_0)^2]$ as a boundary condition. We see that as $x$ increases the amplitude of the retrieved SP pulse is lowered and its width is broadened largely, which means that the quality of the SP memory is low.

The quality of the light memory can be characterized quantitatively by the efficiency and fidelity of the memory.
Taking in consideration of the case where system may have possible gain (e.g. the case in the presence of the microwave field, discussed below), we define the SP memory efficiency as $\eta=1-\eta'$, with
\begin{equation}\label{efficiency}
  \eta'=\frac{|\int^{+\infty}_{-\infty}dt\int^{+\infty}_{0}dz|\textbf{E}^{\textrm{out}}_p(z,t)|^2
  -\int^{+\infty}_{-\infty}dt\int^{+\infty}_{0}dz|\textbf{E}^{\textrm{in}}_p(z,t)|^2|}{\int^{+\infty}_{-\infty}dt\int^{+\infty}_{0}dz|\textbf{E}^{\textrm{in}}_p(z,t)|^2}.
\end{equation}
where $\textbf{E}^{\textrm{in}}_p(z,t)=\textbf{E}_p(0,z,t)$ and $\textbf{E}^{\textrm{out}}_p(z,t)=\textbf{E}_p(L_x,z,t)$.
The fidelity is defined by
\begin{equation}\label{overlap}
  F_s=\frac{|\int^{+\infty}_{-\infty}dt\int^{+\infty}_{0}dz\textbf{E}^{\textrm{out}}_p(z,t)\textbf{E}^{\textrm{in}}_p(z,t+\triangle t)
  |^2}{\int^{+\infty}_{-\infty}dt\int^{+\infty}_{0}dz|\textbf{E}^{\textrm{out}}_p(z,t)|^2\cdot\int^{+\infty}_{-\infty}dt\int^{+\infty}_{0}dz
  |\textbf{E}^{\textrm{in}}_p(z,t+\triangle t)|^2},
\end{equation}
where $\triangle t=36.15\tau_0$ is the time interval between the peak of the input signal pulse $\textbf{E}^{\textrm{in}}_p(z,t)$ and the peak of the retrieved signal pulse $\textbf{E}^{\textrm{out}}_p(z,t)$. Note that the definition of the fidelity (\ref{overlap}) is used to characterize the change of waveshape of the SP during the process of the storage and retrieval of the SP. Based on these formulas, for the case shown in Fig.~\ref{fig3}(c) we obtain $\eta=2.43\%$ and $F_s=69.69\%$. The reasons for so low efficiency and fidelity are  due to the large Ohmic loss in the NIMM and the dephasing in the quantum emitters.

{\large \bf Storage and retrieval of SPs with a microwave field}. For improving the efficiency and fidelity of the SP memory, we apply a weak microwave field~\cite{Eil} to the system to couple the two lower levels of the quantum emitters (i.e. the states $|1\rangle$ and $|2\rangle$), which will provide a gain to the probe field. The switching-on and off of the microwave field can be modeled by
\begin{equation}\label{swith2}
  \Omega_m(x,t)=\Omega_{m}(x)\left[\frac{1}{2}\tanh\left(\frac{t-T_3}{T_s}\right)
  -\frac{1}{2}\tanh\left(\frac{t-T_4}{T_s}\right)\right],
\end{equation}
where $T_s$ is the same as that used in the switch function (\ref{switch1}) of the control field, but
the switching-off and the switch-on times are changed into $T_3$ and $T_4$, respectively. We take
$T_3>T_1$ and $T_4<T_2$, which means that the microwave field plays a role only within the time interval between
the switching-off and the switching-on of the control field [see Fig.~\ref{fig4}(a)], which is necessary to acquire
\begin{figure}
\centering
\includegraphics[scale=0.4]{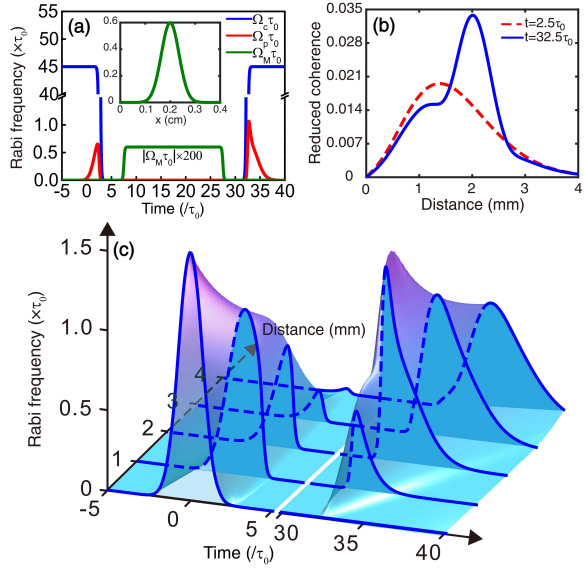}
\caption{\small (color online) SP memory in the presence of the microwave field. (a)~Dimensionless temporal profiles of the control field $\Omega_c\tau_0$ (blue solid line)  and the SP field $\Omega_p\tau_0$ (red solid line), and the microwave field (green solid line) at position  $x=2\, \textrm{mm}$. Inset: the spatial shape of the microwave field. (b)~The coherence of the quantum emitters, $\tilde{\sigma}_{21}$, a as function of distance $x$ at the times $t_1=2.5\, \tau_0$ and $t_2=32.5\,\tau_0$. (c)~The SP  waveshape as a function of $x$ and $t$ during the storage and retrieval in the whole medium.}
\label{fig4}
\end{figure}
a gain for the coherence of the emitters only in the storage period (i.e. $T_1<t<T_2$)  but not to break the EIT condition before and after the storage. In addition, the amplitude of the microwave field is taken to be a function of $x$, which will be used for obtaining high efficiency and fidelity of the SP memory. For a Gaussian-type input of the probe field, one can take
$\Omega_{m}(x)=\Omega_{m0}\exp{-[(x-a)/b]^2}$, here $\Omega_{m0}$ is the amplitude of the microwave field and
$a$, $b$ are two parameters, chosen according to the need for obtaining high memory quality.
Shown in Fig.~\ref{fig4}(a) is the temporal shape of the control field (blue solid line), the SP field (red solid line) and the microwave field (green solid line)  at the position $x=2\,\textrm{mm}$. The inset shows the spatial shape of the microwave field as a function of $x$. When plotting the figure, we have taken $T_3=7.5\tau_0$, $T_4=27.5\tau_0$, $a=2\,\textrm{mm}$, $b=0.5\,\textrm{mm}$, $\Omega_{m0}\tau_0=0.003$ (i.e. $\Omega_{m0}=20\,$kHz) ; other parameters are the same as those used above.

Fig.~\ref{fig4}(b) shows the coherence of the quantum emitters, i.e. $\tilde{\sigma}_{21}$, during the storage period as a function of $x$ at the times $t_1=2.5\, \tau_0$ (red dashed line) and $t_2=32.5\,\tau_0$ (blue solid line). One sees that $\tilde{\sigma}_{21}$ is indeed increased much comparing with the case without the microwave field [Fig.~\ref{fig3}(b)].  Illustrated in Fig.~\ref{fig4}(c) is the SP waveshape as a function of $x$ and $t$ during the storage and retrieval in the whole medium ($L_x=4$\,mm). One sees that as $x$ increases the amplitude (width) of the retrieved SP pulse is larger (smaller) than that without the microwave field. The memory efficiency and fidelity in the presence of the microwave field are given by
\begin{equation}
\eta=10.23\%, \hspace{1cm} F_s=82.21\%,
\end{equation}
improved compared with the case without the use of the microwave field.
If we take $\Omega_{m0}\tau_0=0.009$ (i.e. $\Omega_{m0}=60\textrm{kHz}$), the memory efficiency and the fidelity of are increased into $\eta=98.97\%$ and $F_s=85.62\%$, which means it is feasible to improve the SP memory much by increasing microwave field. A detailed theoretical analysis for the understanding of the principle and for the improvement of the SP memory by the microwave field descried above is given in the Sec.~C of the Supporting information.

\section{CONCLUSIONS}

Our theoretical scheme for the SP memory presented above can be realized in experiment. One of possible examples is to take Pr:YSO as a dopant dielectric and $\textrm{Pr}^{3+}$ ions are doped into a layer near the NIMM-dielectric interface. The levels of the quantum emitters can be chosen as
$|1\rangle=|^3H_4, F=\pm3/2\rangle$, $|2\rangle=|^3H_4, F=\pm1/2\rangle$, and $|3\rangle=|^1D_2, F=\pm3/2\rangle$. Then one has system parameters~\cite{Kuznetsova} $\omega_{31}=3.11\times10^{15}$\,Hz, $\Gamma_2=0.02\,\textrm{Hz}$, $\Gamma_3=12\,\textrm{kHz}$, $\gamma_{21}=10\,\textrm{kHz}$, $\gamma_{31}=\gamma_{32}=9\,\textrm{kHz}$ and $N_a=4.7\times10^{18}\textrm{cm}^{-1}$.
There are many ways to obtain various NIMMs, one of them is a nano-fishnet structure, which has been demonstrated to have a negative-index property in optical frequency region~\cite{Sha,Xiao}. The probe and the control fields can be coupled to the NIMM-dielectric interface by an end-fire coupling with a high coupling efficiency~\cite{Maier,Kam1}. We note that there were some works reporting on the light memory via EIT using solid-state platforms~\cite{Long,England}, but the scheme presented in the present work has many advantages, including: (i)~EIT is largely enhancement by the strong confinement of electric field near the NIMM-dielectric interface, and hence the light field used in the system can be very weak; (ii)~The device fabricated by using such scheme is easy to be miniaturized and thus a large scale integration is possible.

In summary, in this work we have investigated the SP memory via the EIT of the quantum emitters, which are doped at the interface between a dielectric and a NIMM. We have shown that, because of the strong field confinement contributed by the NIMM-dielectric interface, the EIT effect may be largely enhanced. We have also shown that the storage and retrieval of the SP can be realized through the switching-off and switching-on of the control field. In addition, we demonstrated that the efficiency and the fidelity of the SP memory can be improved much by using a weak microwave field. Our analysis can be apply to other interfaces doped with quantum emitters, and the results obtained are helpful not only for enhancing the understanding of SP property but also for practical applications in quantum information processing and transmission.

\begin{suppinfo}

The supporting Information gives detailed deviations of the TM mode of the electromagnetic field, the Maxwell-Bloch equations, and the theoretical analysis for improving the quality of the SP memory by using microwave field.

\end{suppinfo}


\section{Funding Sources}

We acknowledge the support from National Natural Science Foundation of China under Grants No. 11474099 and No. 11475063.

\section{Conflict of Interest}

The authors declare no competing financial interest.


\bibliography{achemso-demo}

\begin{thebibliography}{99}
\bibitem{note100}The reason for a larger (smaller) control-field Rabi frequency obtained in the present system (free atomic gas) is due to the mode confinement (no mode confinenment) in the NIMM-dielectric interface (free atomic gas).
\end{thebibliography}

\end{document}